\documentclass[apj]{emulateapj}
\slugcomment{{\sc Accepted for publication in the Astrophysical Journal Letters} August 12, 2010}
\usepackage{epstopdf}
\usepackage{graphicx}
\usepackage{color}

\shortauthors{El-Mezeini and Ibrahim}
\shorttitle{QPOs in the Reccurent Emission from SGR 1806-20}

\begin{document}

\title{Discovery of Quasi-Periodic Oscillations in the Recurrent Burst Emission \\from SGR 1806-20}
\author{Ahmed M. El-Mezeini\altaffilmark{1,2} and Alaa I. Ibrahim\altaffilmark{1,3,4}}

\altaffiltext{1}{Department of Physics, The American University in Cairo, New Cairo, 11835, Egypt; amezeini@aucegypt.edu, ai@aucegypt.edu}

\altaffiltext{2}{Department of Applied Mathematics \& Theoretical Physics, Centre for Mathematical Sciences, University of Cambridge, CB3 0WA, UK; amiae2@cam.ac.uk}

\altaffiltext{3}{Department of Physics, Faculty of Science, Cairo University, Giza, 12613, Egypt}

\altaffiltext{4}{Fulbright Scholar, Kavli Institute for Astrophysics and Space Research, Massachusetts Institute of Technology, MA 
02139, USA; ai@space.mit.edu}

\begin{abstract}

We present evidence for Quasi-Periodic Oscillations (QPOs) in the recurrent outburst emission from the 
soft gamma repeater SGR 1806-20 using NASA's Rossi X-ray Timing Explorer (RXTE) observations. By searching a sample of 30 bursts for timing signals at the frequencies of the QPOs discovered in the 2004 
December 27 giant flare from the source, we find three QPOs at 84, 103, and 648 Hz in three different 
bursts. The first two QPOs lie within $\sim$ 1$\: \sigma$ from the 92 Hz QPO detected 
in the giant flare. The third QPO lie within $\sim$ 9$\: \sigma$ from the 625 Hz QPO also detected in the same 
flare. The detected QPOs are found in bursts with different durations, morphologies, and brightness, and are 
vindicated by Monte Carlo simulations, which set a lower limit confidence interval $\geq 4.3 \sigma$.  We also find evidence for candidate QPOs at higher frequencies in other bursts with lower statistical significance. The fact that we can find evidence for QPOs in the recurrent bursts at frequencies relatively close to those found in the giant flare is intriguing and can offer insight about the origin of the oscillations. We confront our finding against the available theoretical models and discuss the connection between the QPOs we report and those detected in the giant flares. The implications to the neutron star properties are also discussed. 

\end{abstract}


\section{INTRODUCTION}
Soft Gamma Repeaters (SGRs) and Anomalous X-ray Pulsars (AXPs) are widely believed to be magnetars, i.e. neutron stars with 
intense magnetic field of the order of $10^{14} - 10^{15}$ Gauss \citep{Duncan+Thompson_1992, Woods+Thompson_2006}. Magnetars are characterized by the repetitive emission of bright bursts of low-energy gamma-rays with short durations ($\sim$ 0.1 s) and peak luminosities reaching up to 10$^{41}$ erg s$^{-1}$, well above the classical Eddington luminosity $L_{edd}$ $\approx$ 2$\times$10$^{38}$ erg s$^{-1}$ for a $1.4\, M_\odot$ neutron star. Magnetars are also persistent X-ray pulsars with slow spin periods ($P \sim 2-12$ s) and rapid spin-down rate ($\dot{P}\sim1$ ms$/$yr). Less frequently, magnetars produce enormously energetic giant flares with peak luminosities of the order of 10$^{44}$ - 10$^{46}$ erg s$^{-1}$, followed by pulsating tail lasting for hundreds of seconds. At present, there are 18 magnetars: 8 SGRs and 10 AXPs$\footnote[5]{See the McGill SGR/AXP Online Catalog (http://www.physics.mcgill.ca/$\sim$pulsar/magnetar/main.html)}$ (see  \citet{Mereghetti_2008} for a review). 

The magnetar model postulates that the dominant source of energy of magnetars is their intense magnetic field. The persistent X-ray emission is attributed to magnetospheric currents driven by twists in the magnetic field. The short bursts are produced as a result of outward propagation of Alfv\'en waves through the magnetosphere. The giant flares are triggered by sudden catastrophic reconfiguration of the magnetic field that causes a large scale crustquake  \citep{Thompson+Duncan_1995, Thompson+Duncan_1996, Thompson+Duncan_2001, Schwartz +Zane+Wilson+etal_2005}.

On 2004 December 27, SGR~1806-20 emitted its first recorded giant flare that was observed by a number 
of satellites as the most powerful giant flare from a magnetar. The flare reached a peak luminosity of 
the order of 10$^{46}$ erg s$^{-1}$ in the first 0.2 s \citep{Hurley+Boggs+Smith+etal_2005, 
Palmer+Barthelmy+Gehrels+etal_2005}, causing a strong perturbation in the Earth's ionosphere and 
saturation of most satellite  detectors \citep{Terasawa_2005}. The initial spike was followed by $\sim 400$ s tail modulated 
at the $7.56$ s spin period of the SGR. Using RXTE Proportional Counter 
Array (PCA) observations, \citet{Israel+Belloni+Stella+etal_2005} discovered three QPOs at 18, 30, and 92 Hz in the 
decaying tail. With Ramaty High Energy Solar Spectroscopic Imager (RHESSI), 
\citet{Strohmayer+Watts_2005} confirmed the 18 and 92 Hz QPOs and reported two additional QPOs at 26 
and 626 Hz. Subsequently, \citet{Watts+Strohmayer_2006} re-analyzed the RXTE observations and confirmed 
the 26 and 626 Hz QPOs detected with RHESSI and reported additional QPOs at 150 and 1840 Hz. Two more QPOs 
at 720 and 2384 Hz were reported at a lower significance.

\citet{Strohmayer+Watts_2005} also studied the RXTE observations of the 1998 August 27 giant flare from 
SGR 1900+14 and discovered four QPOs at 28, 53, 84, and 155 Hz in the flare's pulsating tail. In both 
sources the QPOs were detected in part of the tail and were only seen at a specific rotational phase. 
They were identified as a sequence of toroidal modes associated with seismic vibrations in the neutron 
star crust. Other models were also suggested, e.g. \cite{Glampedakis+Samuelsson+Andersson_2006, Levin_2006, Beloborodov+Thompson_2007, Samuelsson+Andersson_2007, Ma+Li+Chen_2008, Shaisultanov+Eichler_2009}.

Motivated by these findings, we undertook the task of searching for similar signatures in the typical 
and more common emission of magnetars: the recurrent bursts. Here we present the results from SGR 1806-20. 
Our results from SGR 1900+14 are forthcoming.

\section{DATA ANALYSIS \& RESULTS}
\label{sec:DataAnalysis}

RXTE first observed SGR 1806-20 during 1996 November 5-18. We obtained the PCA observations in 2--60 
keV and reduced the data with HEAsoft 6.6 to obtain the burst light curves. The observations with the 
most bursts were taken in the event-mode configuration with time resolution $\Delta t \sim 125\mu$s and 
256 energy channels. All 5 proportional counter units (PCUs) of the PCA were fully functional. Photons 
from all PCUs and all detector anodes were combined to maximize the signal-to-noise ratio and the 
artificial event-mode time markers were removed. We screened out very faint bursts that do not have 
sufficient counts and very bright bursts that would cause significant pile-up and dead-time effects and 
selected 30 bursts for our timing analysis. We binned each burst at the smallest time resolution 
allowed by the data configuration ($125~\mu$s) and applied Fast Fourier Transformation (FFT) to the 
binned photon arrival times to obtain a power spectrum for each burst. The frequency resolution 
$\Delta\nu$ = $\frac{1}{\Delta t \times N_{bins}}$ ranged from 1.72 to 14.63 Hz, depending on the burst duration, and the Nyquist 
frequency was $\nu_{nyquist}=\frac{1}{2 \Delta t} \sim$ 4096 Hz. The power spectra were normalized 
using Leahy normalization method \citep{Van_der_Klis_1989, Israel+Stella_1996} and we searched each burst 
power spectrum for timing signals in the frequency range of the QPOs found in the 2004 giant flare. 
Candidate QPOs were fitted to a Lorentzian function to calculate their properties, and their 
significance was investigated using Monte Carlo simulations.

\begin{figure}
\centering
\includegraphics[width=75mm,height=70mm]{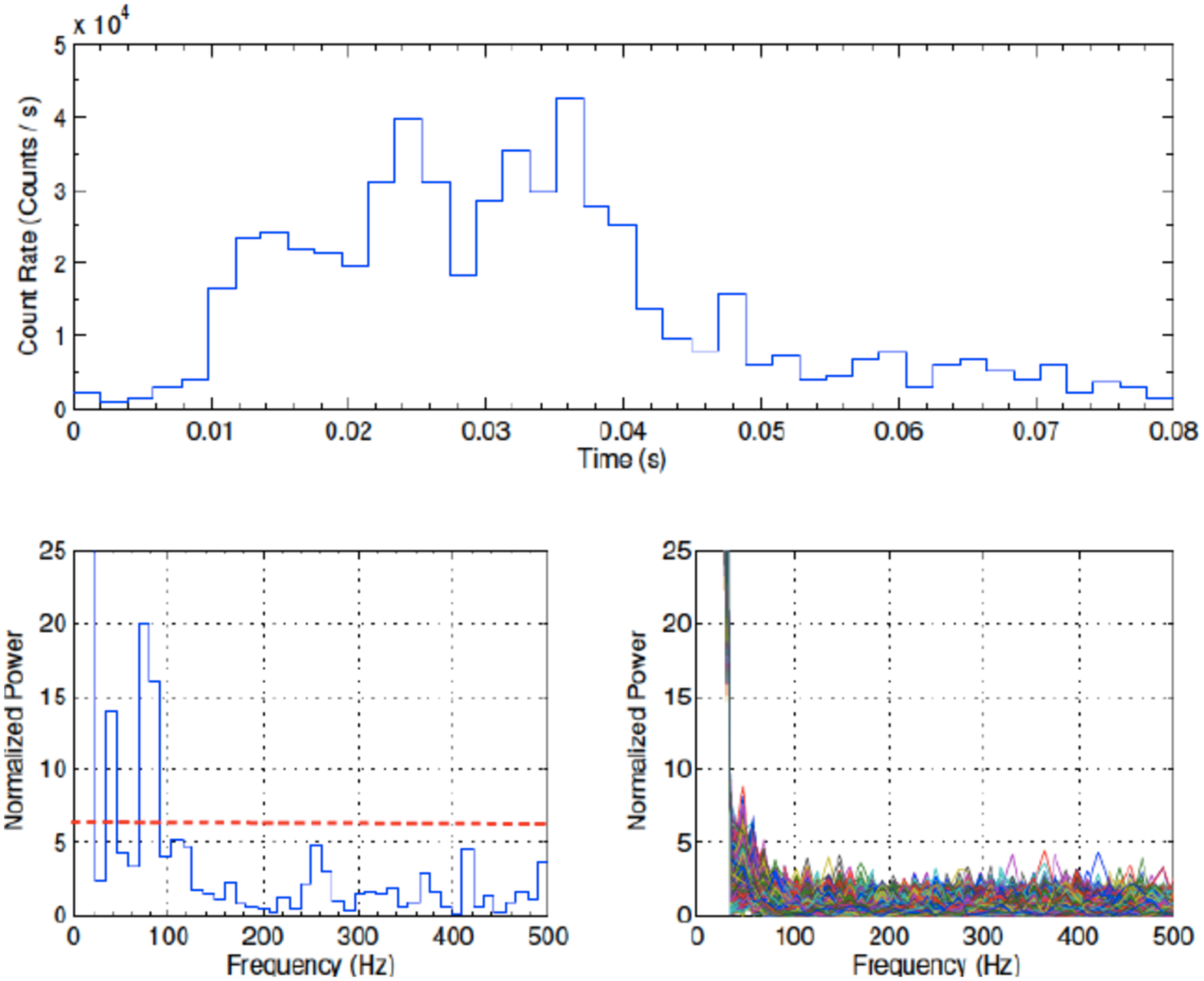}
\includegraphics[width=75mm,height=70mm]{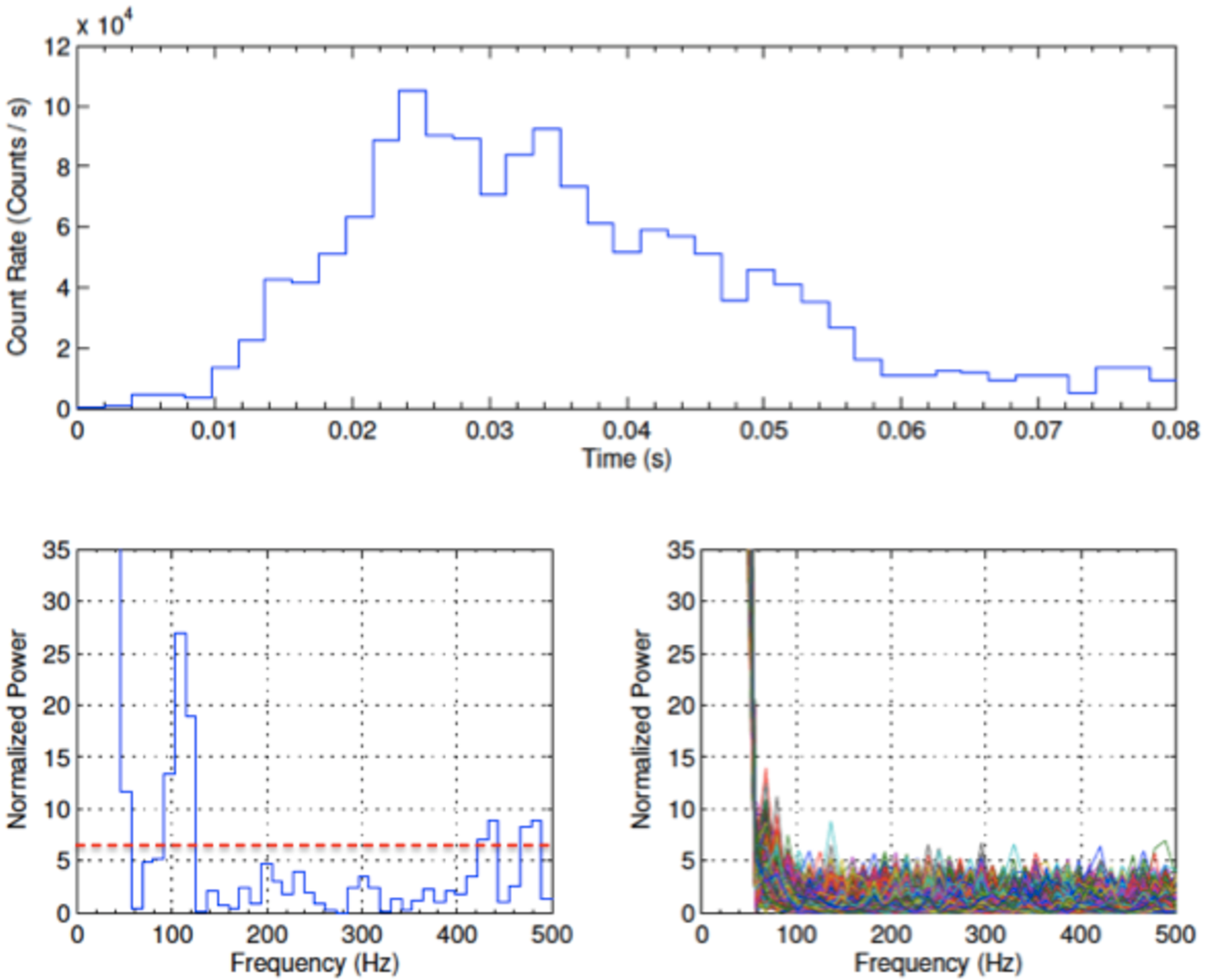}
\includegraphics[width=75mm,height=70mm]{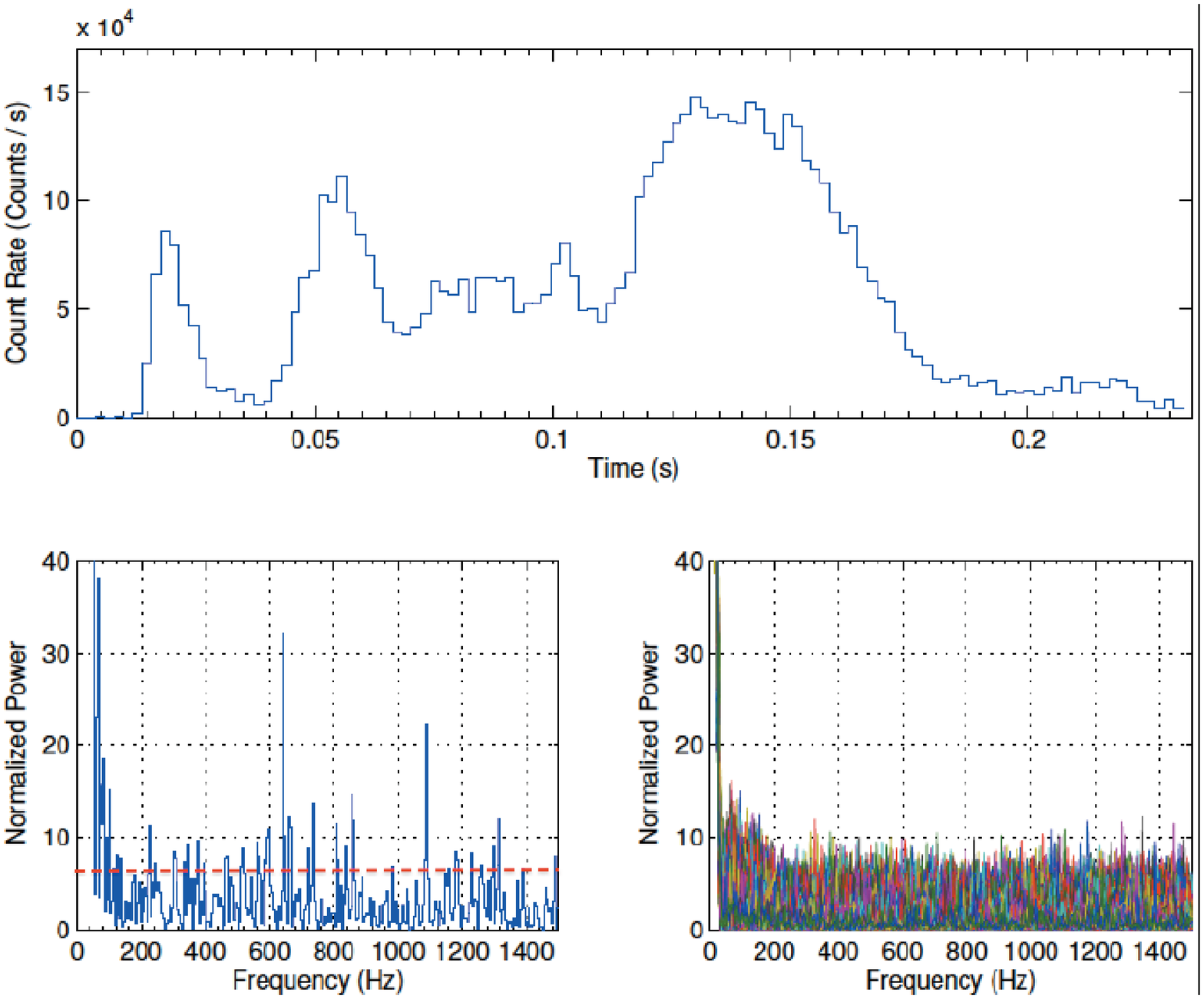}

\caption{Light curves, power spectra, and simulated power spectra of the three bursts that showed evidence for QPOs. Under each burst is its power spectrum (left) and the accumulation of 50,000 power spectra (right) obtained with Monte Carlo simulation. The top burst shows a QPO at 84 Hz ($\Delta\nu$ = 11.38 Hz), the middle burst shows a QPO at 103 Hz ($\Delta\nu$ = 11.38 Hz), and the bottom burst shows a pair of QPOs at 648 and 1096 Hz ($\Delta\nu$ = 4.27 Hz). The dashed horizontal line represents the baseline of the single trial 95$\%$ confidence level. The simulated power spectra reveal no significant structure at or in the vicinity of the detected QPOs. The plotted light curves are binned at 2 ms. The bursts occurred on 1996 Nov. 18 at 06:18:44.446 UTC (top), 08:57:02.083 UTC (middle), and 04:41:53.480 UTC (bottom).}

\label{fig:1}
\end{figure}

In a short ($\sim 0.09$ s), moderately bright burst (total counts $\sim 1.13 \times 10^{3}$), we find a QPO candidate around 84 Hz with single trial 
null probability of 3.4$\times$10$^{-5}$ (see Fig.~1 top panel). The QPO has a centroid frequency $\nu_o = 84.31 \pm 0.66$ 
Hz and a half width at half maximum $\sigma=8.69 \pm 0.78$ Hz, corresponding to a coherence value 
Q $\equiv$ $\nu_o$/$2\sigma$ of 4.84. 

In another short, bright burst with a gradual rise (duration $\sim 0.09$ s and total counts $\sim$ 3.07 $\times 10^{3}$, see Fig.~1 middle panel), we detect a candidate QPO at 103 Hz at a null probability of 1.43$\times$10$^{-6}$. This QPO has a centroid frequency $\nu_o = 103.44 \pm 0.42$ Hz, width $\sigma = 10.65 \pm 0.42$ Hz, and coherence value $Q\sim$ 5. 

In the bottom panel of Fig.~1, we show a long ($>0.1$ s), bright burst (total counts $\sim$  1.37 $\times 10^{4}$) with a pair of QPO candidates at around 648 and 1096 Hz with a single trial probability of 2.33$\times$10$^{-7}$ and 1.37$\times$10$^{-5}$, respectively. Fitting each 
to a Lorentzian function yields $\nu_o= 648.5 \pm 0.15$ Hz, $\sigma= 2.5 \pm 0.09$ Hz, and $Q= 130$ for the first QPO and $\nu_o= 1095.88 \pm 0.21$ Hz, $\sigma= 2.88 \pm 0.17$ Hz, and $Q= 190$ for the second. 

In the other bursts we did not find significant QPOs at the previously detected frequencies. Extending our search to other frequencies, we detect two high frequency QPO candidates in the short, relatively dim burst (duration $\sim$ 0.07 s and total counts $\sim$ 700) shown in Fig.~2 (upper burst) at $1229.8 \pm 0.80$ and $3690.74 \pm 0.67$ Hz. Forming the null hypothesis results in a single trial probability of 3.4$\times 10^{-5}$ for each feature. We estimate that the 1230 and 3690 Hz features have $\sigma$ = 8.00 $\pm$ 0.48 and 8.37 $\pm$ 0.79 Hz respectively and corresponding coherence values of 76 and 220. Another high frequency QPO candidate around 2785 Hz is found in the long, intense burst (duration $\sim$ 0.63 s and total counts $\sim$ 2.00 $\times$ $10^{4}$) shown in Fig.~2 (bottom burst), at a single trial probability of 7.96$\times 10^{-6}$. The QPO is centered at $\nu_o= 2785.36 \pm 0.20$ Hz and has a width $\sigma= 0.75 \pm 0.19$ Hz and coherence $Q= 1860$.


Since our strategy was to search for QPOs at the previously reported frequencies, the number of trials 
for the 84, 103, and 648 Hz QPOs takes into account the number of bursts (30 bursts) and the number of frequency 
bins between the QPOs we detect and those discovered in the giant flare (0.72 bins for the 84 Hz QPO, 0.96 bins for the 103 Hz QPO, and 5.51 bins for the 648 Hz QPO). This brings their null probabilities to 7.34$\times 10^{-4}$, 
4.14$\times 10^{-5}$, and 3.85$\times 10^{-5}$, respectively. For the high frequency QPOs at 1096, 
1230, 2785 and 3690 Hz, we had no prior knowledge of these frequencies and we must account for the 
total number of frequency bins in the power spectra as well as the number of bursts. The number of frequency bins accounted for in the case of the 1096, 1230, 2785 and 3690 Hz QPOs is 960, 280, 2376 and 280 bins respectively. This brings the chance probability to higher than $10^{-3}$, thus ruling out these QPOs on the basis of the null hypothesis alone.


A more cogent approach to assessing the statistical significance is through utilizing Monte 
Carlo simulation. For each of the above bursts, we generated a simulated light curve at the same time 
resolution ($125\,\mu$s) by fitting the burst light curve to a high-order polynomial function. We then 
determined the number of photons in each time bin by seeding a Poisson distributed random number 
generator whose mean value is the number of counts we obtain from the fitting function. We used this 
approach to obtain 50,000 simulated light curves for each real burst and we generated a power spectrum 
for each simulated burst using the same technique we used to find the QPOs in the bursts above. In Fig. 
1 and 2, the right panel under each burst shows the accumulated power spectra of the 50,000 simulated bursts$\footnote[2]{The number of simulated bursts was set by the available CPU and memory 
resources to carry out the Monte-Carlo simulation.}$. We find none to have shown a structure at or near the frequencies of the candidate QPOs with similar or higher power, yielding a null probability lower limit of $\leq$ 2 $\times 10^{-5}$ (or a confidence level interval $\geq 4.3\: \sigma$) for the QPOs reported above. 

The statistical significance arrived at from the two independent approaches above gives a secure 
detection for the 84, 103, and 648 Hz QPOs and a marginal significance for the kHz QPOs.

\begin{figure}
\centering

\includegraphics[width=75mm,height=70mm]{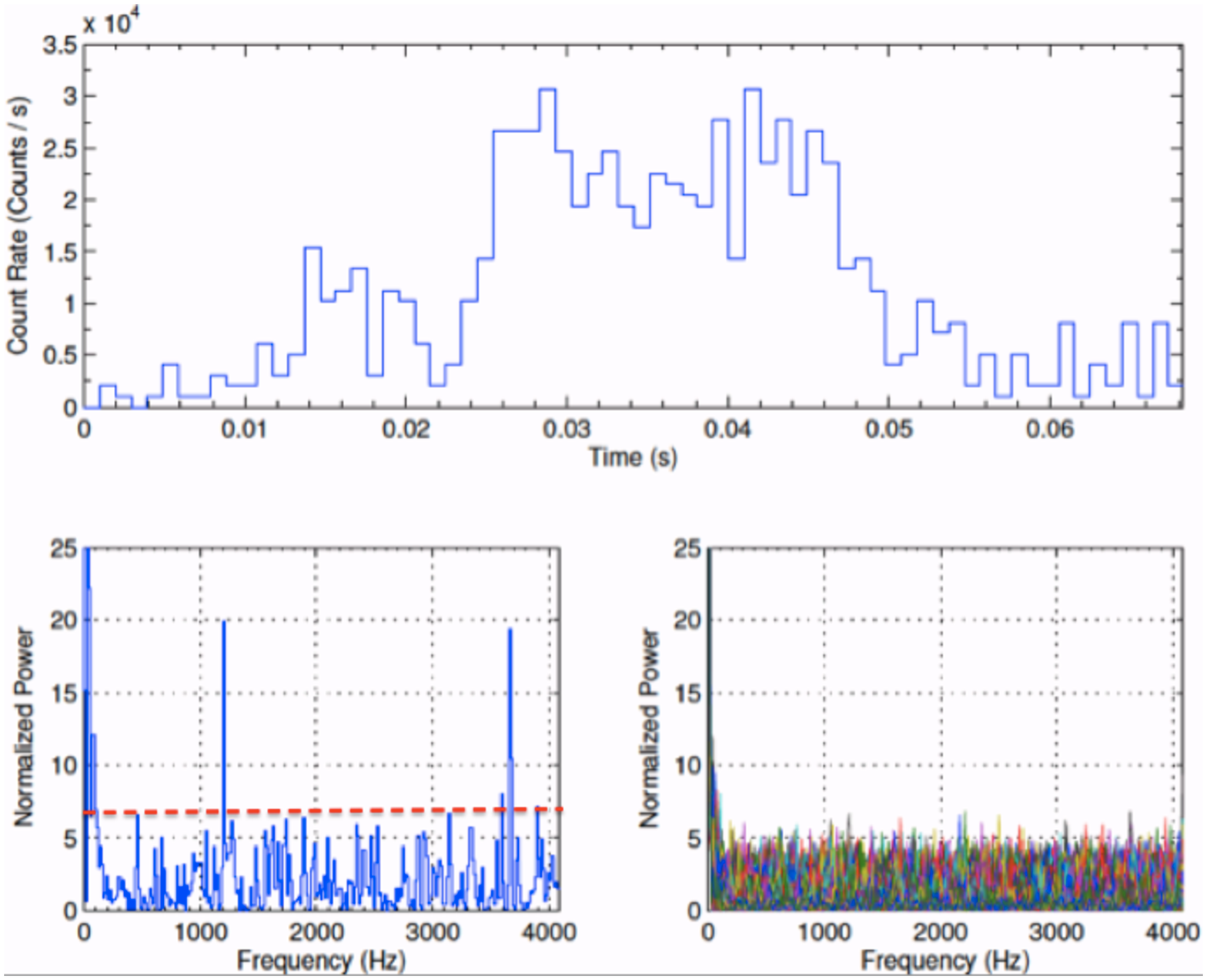}
\includegraphics[width=75mm,height=70mm]{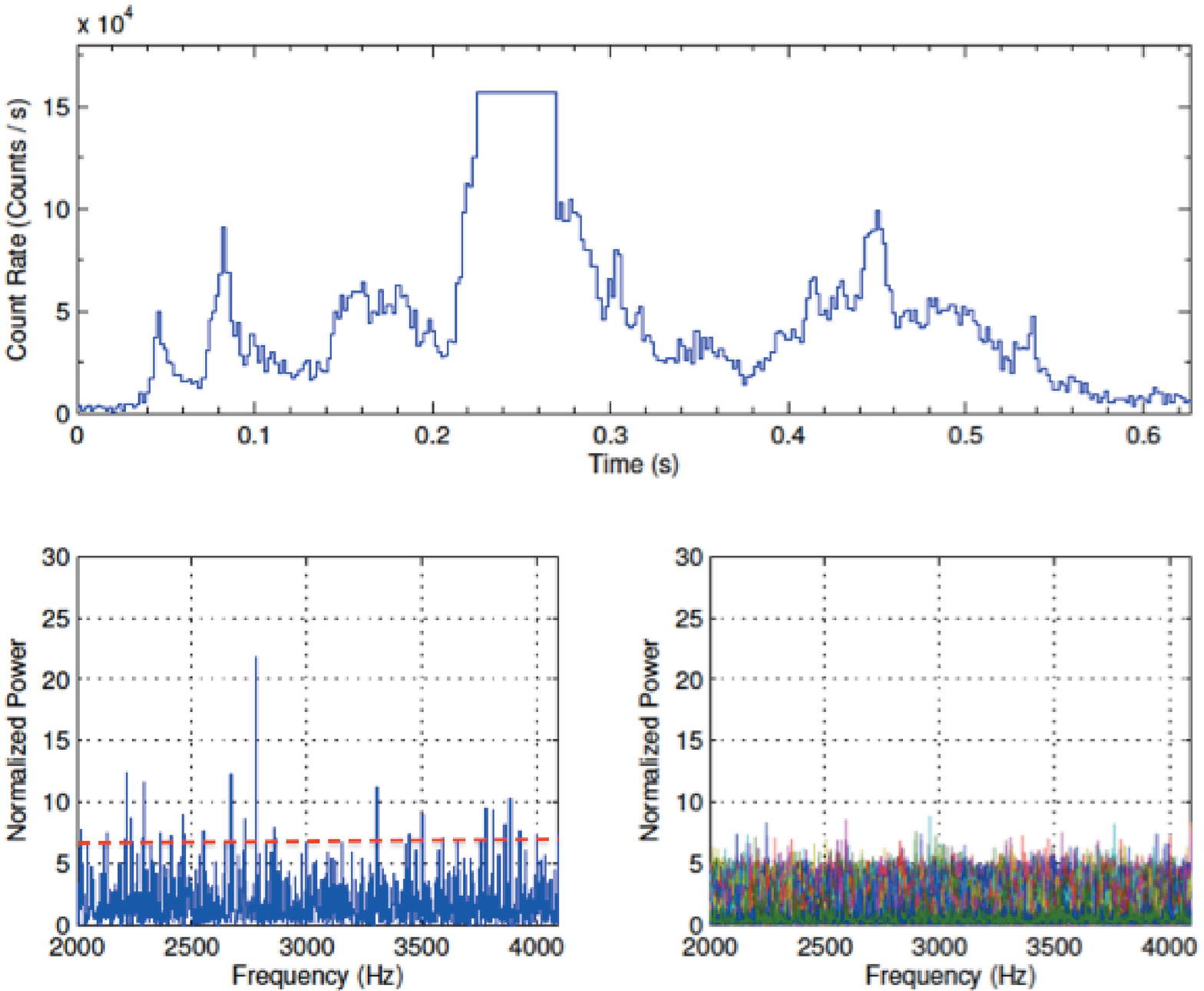}

\caption{Light curves, power spectra, and simulated power spectra of the two bursts that showed evidence for QPOs at high frequencies. Under each burst is its power spectrum (left) and the accumulation of 50,000 power spectra (right) obtained with Monte Carlo simulation. The top burst shows two QPO at $\sim$ 1230 and $\sim$ 3690 Hz ($\Delta\nu$ = 14.63 Hz). The bottom burst shows a QPO at $\sim$ 2785 Hz ($\Delta\nu$ = 1.72 Hz). The  dashed horizontal line represents the baseline of the single trial 95$\%$ confidence level. The simulated power spectra reveal no significant structure at or in the vicinity of the detected QPOs. The plotted light curves are binned at 1 ms (top) and 2 ms (bottom). The bursts occurred on 1996 Nov. 18 at 06:18:44.134 UTC (top) and 12:18:01.487 UTC (bottom).}

\label{fig:2}
\end{figure}


\section{Discussion}


Our search for QPOs in the recurrent bursts from SGR~1806-20 revealed QPOs of interesting similarities to those discovered in the giant flares from the source and from SGR~1900+14 \citep{Israel+Belloni+Stella+etal_2005, Strohmayer+Watts_2005, Watts+Strohmayer_2006}. We found three intriguing QPOs at $84.31 \pm 0.66$, $103.44 \pm 0.42$, and $648.5 \pm 0.15$ Hz. The 84 Hz QPO lies within $\sim$ 1$\: \sigma$ from the $92.5 \pm 0.2$ Hz QPO from the same source and coincides with the 84 Hz QPO discovered in the SGR 1900+14 giant flare. The 103 Hz feature lies within $\sim$ 1$\: \sigma$ from the $92.5 \pm 0.2$ Hz feature from the source. The high frequency QPO at $648 \pm 0.15$ Hz is within $\sim$ 9$\: \sigma$ from the $625.5 \pm 0.15$ and $626.5 \pm 0.02$ Hz QPOs detected from the source with RXTE and RHESSI, respectively. Those QPOs are detected in bursts with different duration, luminosity, rise time, and light curve morphology, and thus cannot be due to artifacts in the light curves. Their statistical significance is further confirmed by the Monte-Carlo simulations.  

Torsional shear mode mechanism on the neutron star crust is a promising framework for interpreting magnetar QPOs since the observed low frequencies fit comfortably within theoretical models predictions and are associated with a sequence of $\it{l}$ modes, while the high frequency QPOs are interpreted as modes with higher order radial displacement eigenfunction ($\it{n}\geq$ 1). As a consequence of the twisting of the magnetic field lines and the outward propagation of the Alfv\'{e}n waves, stresses build up in the neutron star crust, which is very likely to excite toroidal vibrational modes in the neutron star crust \citep{Ruderman_1991,Ruderman+Zhu+Chen_1998,Schwartz +Zane+Wilson+etal_2005}. The excited frequency modes depend on the neutron star mass and radius, crustal composition and the magnetic field \citep{Duncan_1998, Hansen+Cioffi_1980, McDermott+van Horn+Hansen_1988, Messios+Papadopoulous+Stergioulas_2001, Piro_2005, Schumaker+Thorne_1983, Strohmayer_1991}. 


In the context of global seismic vibrations, \citet{Duncan_1998} estimated the eigenfrequency of the fundamental toroidal mode ${\it l}$=2, ${\it n}$=0 denoted $_{2}\nu_o$ to be

\begin{equation}
\nu(_{\it 2}{\it t}_{\it o}) = 29.8 \frac{\sqrt{1.71-0.7M_{1.4}R_{10}^{-2}}}
{{0.87R_{10}+0.13M_{1.4}R_{10}^{-1}}} ~\textnormal{Hz}.
\end{equation}

Where $R_{10}$ $\equiv$ $R/10$ km and $M_{1.4} \equiv M/1.4M_\odot$. If a twisted magnetic field is embedded throughout the neutron star crust, one needs to account for the magnetic field correction and the eigenfrequencies are modified to $\nu=\nu_o\left[1+\left(\frac{B}{B_\mu}\right)^2\right]^{1/2}$ Hz, where $\nu_o$ is the non-magnetic eigenfrequency, $B_{\mu}$ $\equiv$ $\sqrt{4\pi\mu}$ $\approx$ 4$\times$10$^{15}$ $\rho_{14}^{0.4}$ Gauss and $\rho_{14}\approx$1 is the density in the deep crust in units of 10$^{14}$ g cm$^{-3}$. The higher order mode eigenfrequencies ($\it{n}$=0 \& $\it{l}>$2) were shown to be scaled as $\nu(_{\it l}{\it t}_{\it o})$ $\propto$ $[{\it l}({\it l}+1)]^{1/2}$ \citep{Hansen+Cioffi_1980} and were derived by \citet{Strohmayer+Watts_2005} under the assumption that the magnetic tension boosts the field isotropically. One must note that the degree to which the magnetic field modifies the eigenfrequency is highly dependent on the magnetic field configuration and other non-isotropic effects which could alter this correction significantly \citep{Messios+Papadopoulous+Stergioulas_2001}. The eigenfrequencies of the higher order toroidal modes are given by

\begin{equation}
\nu(_{\it l}{\it t}_{\it o})=\nu(_{\it 2}{\it t}_{\it o})\left[\frac{{\it{l}(\it{l}+\textnormal{1})}}{6}\right]^{1/2}
\left[1+\left(\frac{B}{B_\mu}\right)^2\right]^{1/2} \textnormal{Hz}
\end{equation}


The QPOs at 92.5 and 30 Hz discovered in SGR 1806-20 giant flare are attributed to the $\nu(_{\it 7}{\it t}_{\it o})$ and $\nu(_{\it 2} {\it t}_{\it o})$ modes, respectively while the 626.5 Hz QPO was modeled as ${\it n}$=1 toroidal mode where the choice of $\it{l}$ is irrelevant. In this context, the 84 and 103 Hz QPOs that we report correspond to $\nu(_{\it 6}{\it t}_{\it o})$ and $\nu(_{\it 8}{\it t}_{\it o})$  respectively. It was recently suggested that a number of peaks can be seen in the in the 78Ð105 Hz frequency range of the power spectrum of SGR 1806-20 giant flare that could be attributed to magnetic splitting under the perturbative condition $\frac{B^2}{8\pi\mu} \ll 1$ \citep{Shaisultanov+Eichler_2009}. In that view, the two QPOs at 84 and 103 Hz may be the result of magnetic splitting of a degenerate mode.

The 648 Hz QPO we report may be interpreted in terms of ${\it n}$=1 toroidal mode irrelevant of the choice of $\it{l}$. For such high frequency QPOs, the ${\it n}\ge$1 torsional modes in the neutron star crust is a more plausible interpretation and they fall within the frequency range suggested for the search for higher frequency QPOs by \citet{Piro_2005}. 

The detection of QPOs in magnetars has triggered a boost in the theoretical modeling that included attempts to construct more refined models that consider crust/core coupling, magnetic field geometry, elastic properties of the crust and other effects. \citet{Glampedakis+Samuelsson+Andersson_2006} used a simple global MHD toy model to calculate the global magneto-elastic modes taking into consideration the magnetic coupling between the elastic crust and the fluid core under the assumption that the magnetic field is uniform. The model provides intriguing results and indicates that the system has a rich spectrum of global modes and it will naturally excite modes that minimize the energy transfer between the crust and the core. The excited modes frequencies are similar to those of the observed QPOs, where the modes that survive are those for which the coupling is minimum and have very weak amplitudes in the core. This model further supports the global seismic vibrations interpretation of the observed QPOs and it also accommodates for the 18 and 26 Hz QPOs seen in the SGR~1806--20 giant flare. A general relativistic formulation using the Cowling approximation, which does not include any correction for the magnetic field yielded a set of analytic estimates of the eigenfrequency modes based on numerical calculations \citep{Samuelsson+Andersson_2007}. This model yields a different scaling with $\it{l}$ than the previous $[{\it l}({\it l}+1)]^{1/2}$ scaling, which leads to different mode assignments. Futhermore, the assumption of an isotropic magnetic field is inadequate as recent MHD simulations of the evolution of the magnetic field and its geometry in magnetars have indicated that magnetars have a strong dipole field component and a possible strong internal poloidal field \citep{Braithwaite+Spruit_2006}. Since the QPOs we report and those observed in the giant flare are about 8 years apart, the observed frequencies can offer insight on the magnetic field decay of the neutron star. If we assume that the 103 and 92 Hz QPOs are attributed to the $\it{l}$=7 mode, we infer that the magnetic field strength of SGR 1806-20 has decayed from 2$\times10^{15}$ in 1996 to 3.5$\times10^{14}$ G in 2004. 

Magnetospheric oscillation models were suggested as a possible interpretation for the origin of magnetar QPOs. \citet{Ma+Li+Chen_2008} argued that the low frequency range QPOs could be modeled using the standing sausage mode oscillations of flux tubes in the magnetar corona. It is assumed that part of the plasma ejected during the giant flare is trapped by the magnetic field and then forms magnetic flux tube structures similar to what is seen in the solar corona. The tube oscillation model cannot accommodate the high frequency QPOs in SGR 1806-20. For the very high frequency QPOs, \citet{Beloborodov+Thompson_2007} suggested that the inner electric current in magnetar corona will induce very high frequency QPOs $\sim$ 10 kHz where the corona self organizes itself into quasi steady state. Such high frequency QPOs remain to be detected while the lower frequency oscillations would involve the outer corona where the emission in this region is not intense enough to excite the observed QPOs. The QPOs we report are detected in short ($<0.1$ s), relatively dim bursts and in long ($>0.1$ s), bright bursts that would produce different signatures in the magnetosphere. Since the short, weak bursts may not produce a fireball at all or be of magnetospheric origin, this casts doubts on magnetospheric models as an origin for the QPOs in magnetars. 

A number of theoretical models postulated that AXPs and magnetars are neutron stars surrounded by fossil disks that were acquired during supernova collapse or during a common-envelope interaction \citep{Chatterjee+Hernquist+Narayan_2000, Chatterjee+Hernquist_2000, Perna+Hernquist+Narayan_2000}.
\citet{Wang+Chakrabarty+Kaplan_2006} reported evidence for a debris disk around the AXP 4U 0142+61, which led to the suggestion that the magnetar QPOs could be due to a possible disk as in the case of accreting neutron stars, which exhibit kHz QPOs that are believed to originate in the accretion disk \citep{Van_der_Klis_2006}. The reported QPOs are detected during the bright X-ray emission of the bursts and are not seen in the decaying phase, thus arguing against arising from a possible passive debris disk around the magnetar. Furthermore, this interpretation is mainly ruled out by the fact that we have no evidence for the existence of a fallback debris disk around SGRs. \citet{Watts+Strohmayer_2007} pointed out that for neutron stars that exhibit extremely high frequency QPOs, the inner disk radius of the debris disk is comparable to that of the neutron star. In the case of magnetars, the inner disk radius would have to be several solar radii to induce the observed magnetar QPOs, which rules out the modes of a debris disk mechanism as a viable interpretation for the magnetar QPOs.

Levin (2006) argued against the global seismic vibrations as a viable interpretation for the QPOs in magnetars and suggested that the QPOs could be associated with MHD elastic modes of the neutron star. However, the toy model employed by Levin assumed a rather simplified model for the geometry of the neutron star and the boundary conditions. \citet{Watts+Strohmayer_2007} offered a detailed critique and casted doubts about Levin's conclusions except for the crust/core coupling effect which needs to be well accounted for in future models.




Observing QPOs in both the giant flares and the recurrent burst emission from magnetars is an important development in our understanding of neutron star physics. Further observations, analysis of archival data, and model developments are key to shedding further insight on the underlying phenomena.

\section{ACKNOWLEDGEMENTS}

We would like to thank Anna Watts and Philip Kaaret for their insightful comments and Gianluca Israel and Craig Markwardt for the useful suggestions. A.M. is grateful to Kent Wood and Christopher Thompson for advice and useful discussions during the Neutron Stars and Gamma Ray Bursts Conference that was held in Cairo and Alexandria, Egypt in 2009. We acknowledge the MIT Kavli Institute of Astrophysics and Space Research where this work was partly carried out. A.M is funded by AUC research grant. A.I. acknowledges support from the Fulbright Commission. 

\end{document}